\documentclass[draft]{agujournal2018}
\usepackage{natbib}
\usepackage{apacite}
\usepackage{url} 

\usepackage{setspace}

\journalname{JGR-Planets}
\begin{document}

\title{The Pressure and Temperature Limits of Likely Rocky Exoplanets}

\authors{C.T. Unterborn\affil{1} and W.R. Panero\affil{2}}
\affiliation{1}{School of Earth and Space Exploration, Arizona State University}
\affiliation{2}{School of Earth Sciences, The Ohio State University}

\correspondingauthor{C.T. Unterborn}{cayman.unterborn@asu.edu}

\begin{keypoints}
\item Most rocky exoplanets are no more than 50\% greater in radius and no more than 5 times Earth's mass.
\item Uncertainties in mineral equations of state have a minimal effect on modeled planetary radius and inferred interior properties.

\item The maximum likely pressure of silicate mantles in super Earths is $\sim$630 GPa for all but those planets with FeO-rich compositions.
\end{keypoints}

\begin{abstract}
The interior composition of exoplanets is not observable, limiting our direct knowledge of their structure, composition, and dynamics. Recently described observational trends suggest that rocky exoplanets, that is, planets without significant volatile envelopes, are likely limited to $<$1.5 Earth radii. We show that given this likely upper limit in the radii of purely-rocky super-Earth exoplanets, the maximum expected core-mantle boundary pressure and adiabatic temperature is relatively moderate, 630 GPa and 5000 K, while the maximum central core pressure varies between 1.5 and 2.5 TPa. We further find that for planets with radii less than 1.5 Earth radii, core-mantle boundary pressure and adiabatic temperature are mostly a function of planet radius and insensitive to planet structure. The pressures and temperatures of rocky exoplanet interiors, then, are less than those explored in recent shock-compression experiments, ab-initio calculations, and planetary dynamical studies. We further show that the extrapolation of relevant equations of state does not introduce significant uncertainties in the structural models of these planets. Mass-radius models are more sensitive to bulk composition than any uncertainty in the equation of state, even when extrapolated to TPa pressures.

\end{abstract}
\section{Introduction}
Of the $\sim$4000 exoplanets known today (\url{https://exoplanetarchive.ipac.caltech.edu}), the most common is a planet unlike anything in the Solar System: larger than the Earth and smaller than Neptune \citep{Peti13}. These planets are described as being ``super Earths'' or ``mini-Neptunes,'' where a ``super Earth'' is described as a primarily rocky planet without a significant gaseous envelope, while a ``mini Neptune'' is dominated by a gaseous envelope of either H$_2$/He or water vapor. The description of the nature, structure, and evolution of such planets therefore is hindered by lack of a likely analogue in the Solar System. Our Solar System contains four rocky planets, each with unique surface, atmosphere, and dynamical history. Of these, the Earth stands alone, not just in the habitable state of its surface and atmosphere, but also in its interior dynamics, expressed at the surface as plate tectonics. These dynamics are a complex function of bulk chemical composition, chemical differentiation, and formation history. 

Extrasolar rocky planets will also likely display a wide variety of surface characteristics and interior compositions. These planets represent a new frontier in geoscience: one where most constraints of planetary composition and structure need to be relaxed. In our search for ``Earth-like'' planets we must therefore characterize, however broadly, their geochemical and geodynamical states. This requires knowledge of the physical properties of materials under the relevant pressures and temperatures, which, in turn, is a function of the planet's bulk composition and mass.  Insight into the composition of an exoplanet is most frequently gained through measurement of both mass and radius of the planets (the ``mass-radius relationship'') from which mean density can be calculated and a planet's interior composition might be inferred \citep[e.g. ][]{Vale06,Seag07}. However, planetary mass is rarely measurable to precision better than 20\%, and planetary mean density is degenerate with respect to relative proportions of metallic core, silicate rock, and H$_2$/He atmosphere or H$_2$O layer \citep{Dorn15,Unte18a,Unte18b}. 

\begin{figure}[t!]  
    \centering
    \includegraphics[width=.8\linewidth]{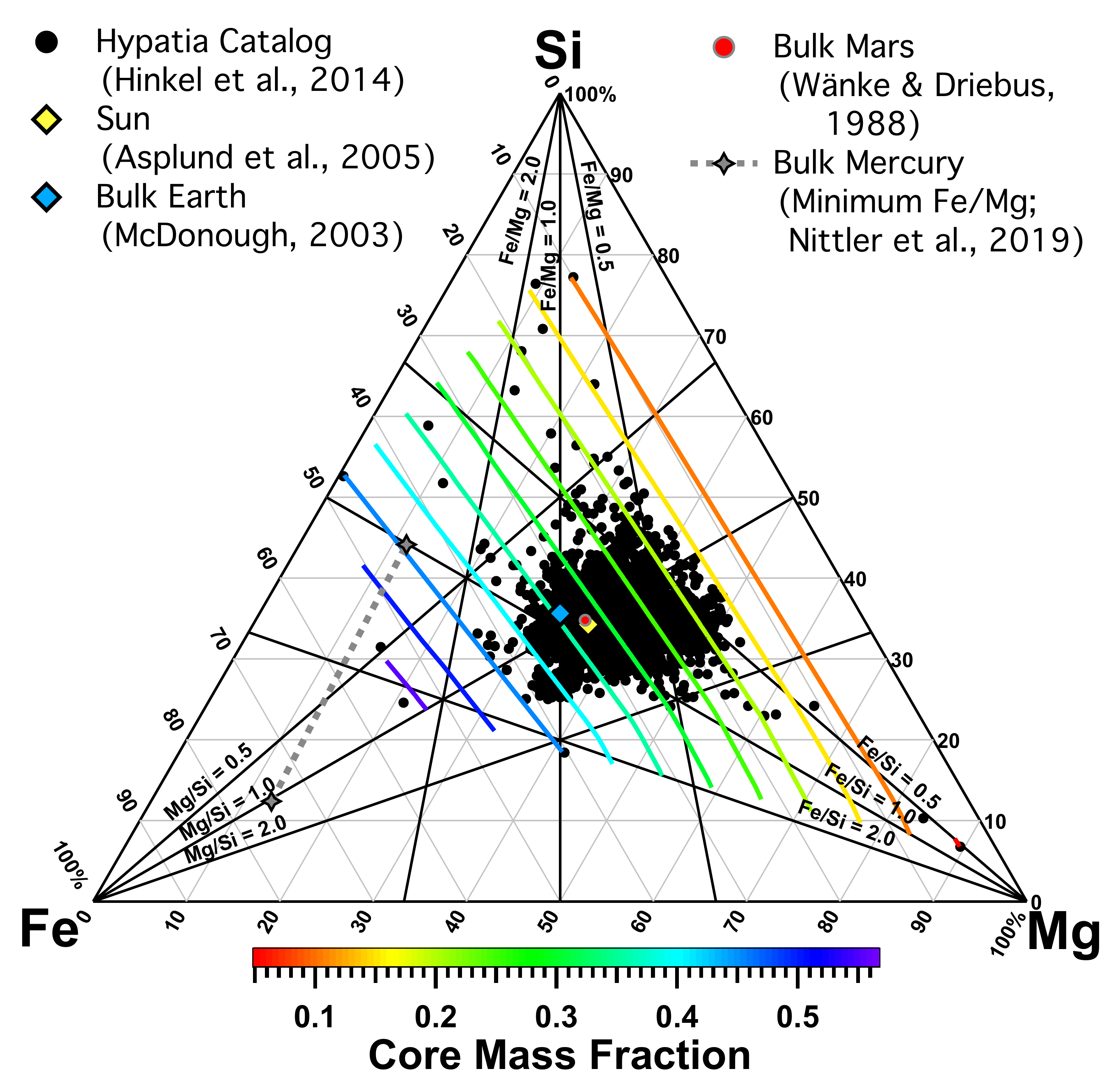}
    \caption{Ternary of molar abundances of Fe, Mg and Si for a sample of 3300 FGK stars as normalized in the Hypatia Catalog \citep{Hink14}. Colored contours represent the calculated CMF assuming a silicate mantle with stellar Si/Mg ratio and a CMF that is the mass ratio of Fe/(MgO+SiO$_2$). For reference we include the Solar composition \citep[yellow diamond, ][]{Aspl05}, the bulk Earth \citep[blue diamond, ][]{McD03}, bulk Mars \citep[red circle, ][]{wanke94} and the range of possible bulk Mercury compositions assuming no FeS layer is present  \citep[gray dashed line with stars, ][]{Nitt17}.}

    \label{fig:Tern}
\end{figure}

In order to account for this degeneracy on interior composition, the relative proportions of the dominant rocky planet-building elements (Mg, Si, Fe) of the host star may be adopted as a proxy for bulk planet. These elements are the most abundant refractory elements, with similar condensation temperatures within the protoplanetary disk \citep{Lodd03}, and are therefore not expected to chemically fractionate relative to each other during formation. Large catalogs of the elemental abundances \citep[e.g. the Hypatia Catalog: ][]{Hink14} in the atmospheres of Sun-like stars (e.g. molar Fe/Mg, Si/Mg), demonstrate that these relative abundances can vary by as much as factors of 2 (Figure \ref{fig:Tern}).

A simple metric of this expected diversity in interior structure is how the relative ratios of Fe, Si and Mg affect the estimates for the core mass fraction (CMF) of a planet. Together with oxygen, these 4 elements account for 95\% by mole of all elements in the Earth \citep{McD03}. Assuming a silicate mantle with stellar Si/Mg ratio, and all Fe resides in the core, CMF is simply the mass ratio Fe/(MgO+SiO$_2$). Adopting compositions of 3300 FGK-type stars (the Sun is a G-type star) from the Hypatia Catalog \citep{Hink14}, we calculate a CMF range of $0.05 < \rm{CMF} < 0.55$ for the entire catalog (Figure \ref{fig:Tern}). A solar composition predicts an associated rocky planet with a CMF $\sim0.33$, similar to Earth's CMF of 0.323. Comparable ranges of abundances are present for other geochemically and geophysically important elements such as Al, Ca, Na and Th \citep{Hink18, Unter15}. If stellar chemical diversity manifests as rocky planet chemical diversity, then the likely range in potential rocky exoplanet composition includes those compositions sufficiently different than those found in the Solar System. As such, many of these compositions are mostly unexplored experimentally, limiting our ability to self consistently model the geochemical and geodynamical consequences of these novel planetary makeups. 

Compounding the problem of quantifying the composition of a planet are the inherent uncertainties in the measurement of mass and radius \citep{Gras09}, measurement of stellar abundances \citep{Hink18} and those within the underlying mass-radius model \citep{Dorn15, Unte16}. One oft-overlooked source of uncertainty when calculating planetary interior models is the uncertainty in the equation of state of the planetary materials. Equations of state for relevant rocky planetary materials are often measured at pressures and temperatures well below those expected in super-Earth's and mini-Neptunes, and instead the resulting data is extrapolated well beyond the experimental conditions \citep{Unte16,Smith18}. For a 1$R_\oplus$ (1 Earth radius = 1$R_\oplus$ = 6371 km) planet, \citet{Unte16} demonstrated that differences between adopted equations of state of $\epsilon$-Fe has a 3\% effect on the predicted planet's mass assuming an Earth-like composition. In contrast, \citet{Smith18} showed that the uncertainty of equation of state has an increasing impact on the inferred mass of a planet, with the uncertainty of the iron core density in excess of 15\% at 5 $M\oplus$ (1 Earth mass = 1$M_\oplus = 5.972\times10^{24}$ kg). Taken together, these uncertainties may make it difficult to distinguish super-Earths from mini-Neptunes.

Over its 9 year mission, the Kepler survey discovered more than 2300 planets by detecting the eclipse of planets passing in front of their host star. This large dataset therefore allows us to explore exoplanet demographics and population statistics as a function of planetary radius, while, for a subset of these planet's, the mass is measurable through subsequent ground-based radial velocity surveys. Analysis of the occurrence rates of exoplanets as a function of radius alone, however, revealed new insights into the architecture of planets, with clues to their formation and evolution \citep{wolf15}. Most significantly, the Kepler survey revealed that super-Earths and mini-Neptunes are the most abundant planets in the Galaxy \citep{Peti13,Pasc18}. A closer look at the Kepler sample shows that the distribution can be grouped into two distinguishable radial populations: those less than 1.5 Earth radii $R_\oplus$ and those between 2.0-4 $R_\oplus$, with a relative minority of planets in the radius range of 1.5-2.0 $R_\oplus$(\url{https://exoplanetarchive.ipac.caltech.edu}). This ``radius-gap'' in the occurrence rate of super-Earths and mini-Neptunes has been confirmed when statistical corrections to insure completeness within the Kepler sample \citep{Fult17}. The ``radius gap'' is currently only robustly confirmed for planets with orbits $<100$ days, where the exoplanet demographic data set is most complete. Population data also hint that the phenomenon extends well beyond this orbital period cutoff to planets with periods between 300-700 days \citep{Burke15}. Whether the cause of the gap is due to photoevaporation of a primordial atmosphere for planets that formed close-in to their host star \citep[e.g.][]{Lope16}, or the late migration of volatile-rich planets from outside the snowline to close-in orbit \citep[e.g. ][]{Raym08}, is unknown and an active area of astrophysical research.

Analysis of the average density of planets within Kepler's different radial bins below 4 $R_{\oplus}$ \citep[e.g. ][]{Weis14,Roge15}, shows a linear increase of density with increasing planetary radius and transitioning to an exponentially decreasing density with radius. Combining occurrence rate data of the ``radius-gap'' with planetary formation and evolution models leads to the growing consensus that these trends represent two classes of planets: those primarily made of iron cores and rocky mantles, with minimal volatile-rich atmospheres, and  another containing significant low-density volatile layers. These two groups of planets are interpreted being high-density super-Earths and volatile-rich mini-Neptunes, respectively. This transition between super-Earth and mini-Neptune is observed to occur at the low end of the radius gap, at 1.5 $R_\oplus$ \citep{Weis14,Roge15}. That is not to say that purely rocky planets of radii greater than 1.5 $R_\oplus$ do not exist, rather that planets with greater radius are more likely to have a significant gaseous envelope and thus be mini-Neptunes.

Here we model the interior structures of these most likely-to-occur rocky planets, whose radius are less than 1.5 $R_\oplus$. We consider the impact of the uncertainty in the thermoelastic parameters of the planetary materials at high pressures and temperatures. We note that the relevant pressures and temperatures of these planetary interiors often exceed the accessible range of experimental methods.  We show that the planet's central pressure, core-mantle boundary pressure and adiabatic temperature gradient are primarily dependent on the total planet radius and secondarily on the relative mass fraction of the core. We also compare our fully self-consistent model with those of simplified empirical models \citep[e.g. ][]{Zeng16, Zeng17}.

\section{Methods}
Models of planet radius, $R$, as a function of planet mass, $M$, are performed using the ExoPlex mass-radius software package \citep{Unte18a} as a function of bulk planet composition (Mg, Fe, Si, O) and mantle potential temperature ($T_{Pot}$). ExoPlex determines the radius, radial density, mantle temperature, pressure and gravity as a function of bulk planetary composition and total mass. For the purposes of these models, we simplify those aspects of bulk mantle chemistry that have a negligible effect on a planet's mean density, such as the inclusion of Ca and Al \citep{Dorn15,Unte16}. We therefore adopt a simplified two layer planet of a silicate mantle and solid metallic (Fe) core. The silicate mantle composition is assumed to have a fixed molar ratio Si/Mg = 1. This value is chosen for convenience, however, as changes in relative core mass and core chemistry are the dominant compositional controls that affect the results of mass-radius-composition calculations \citep{Dorn15, Unte16}.

\begin{figure}[t!]
    \centering
    \includegraphics[width=0.6\linewidth]{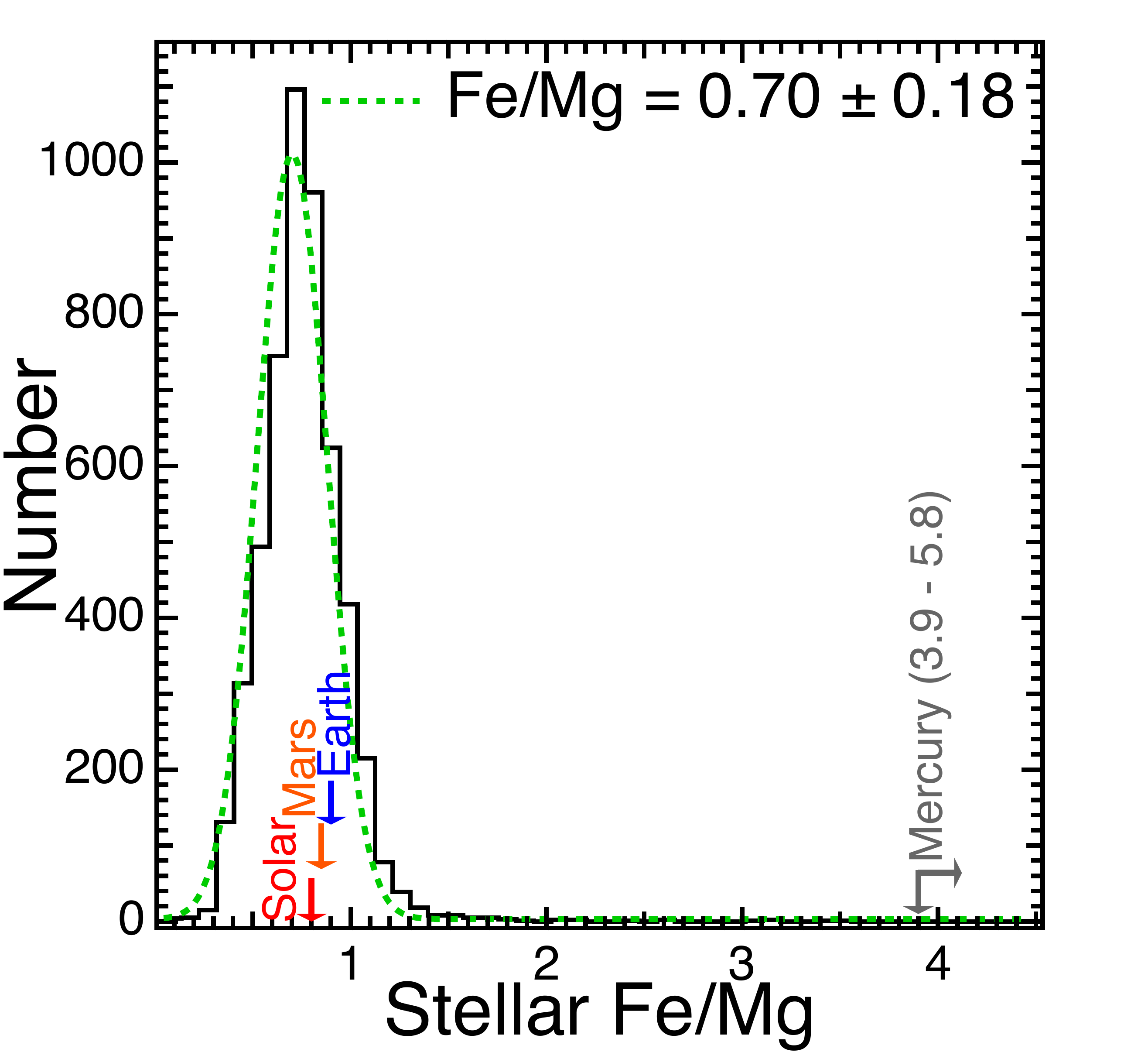}
    \caption{Histogram of Fe/Mg for 5220 FGK stars taken from the Hypatia catalog \citep{Hink14}. The Solar \citep{Lodd10}, bulk Earth \citep{McD03}, bulk Mars \citep{wanke94} values are shown in red, blue, and orange, respectively, and the range for bulk Mercury ($3.9 \leq$ Fe/Mg $\leq5.8$) assuming no FeS layer is present \citep{Nitt17} is shown in gray. }
    \label{fig:FeMg_hist}
\end{figure}

We focused our analysis on planetary compositions that reflect their stellar compositions, and therefore model a range of core mass fraction (CMF), as determined using FKG-star compositional extremes. For our simple stoichiometric approach, a planet's CMF is a function of its bulk Fe/Mg. The average molar Fe/Mg for FGK stars is 0.7$\pm$0.18 \citep{Hink14} (Figure \ref{fig:FeMg_hist}). For our calculations, we adopt two end-member Fe/Mg values for our models of 0.6 and 1.5, representing 80\% of the observed stellar compositions (Figure \ref{fig:FeMg_hist}). This includes the implicit assumption that refractory composition of a planet reflects the refractory composition of the star. For this stoichiometry, our model values reflect CMF between 0.25 and 0.45 and mantle mineralogy dominated by olivine, pyroxenes, Mg-perovskite (Mg-Pv), and post-perovskite (Ppv), with these oxides disassociating into MgO and SiO$_2$ at high pressure and temperature (Section \ref{sec:MRD}). 

The range of Fe/Mg observed in our own Solar System is within the range explored in our models for the majority of the terrestrial planets (Figure \ref{fig:FeMg_hist}). The Solar value of Fe/Mg = 0.8 \citep{Lodd03} falls within 10\% of those derived from geochemical models of the bulk Earth \citep[Fe/Mg = 0.9; ][]{McD03} and bulk Mars \citep[Fe/Mg = 0.85; ][]{wanke94} with a poorly constrained Venusian value remaining unknown but consistent with the Earth value \citep{Zhar83}. This similarity in bulk Fe/Mg is despite the fact that Mars contains 10 wt\% greater FeO than the Earth's mantle \citep{wanke94}. In contrast, Mercury has a molar iron to magnesium ratio significantly greater than the Sun and Earth, between 3.9 and 5.8 \citep{Nitt17}. The exact value of Fe/Mg is dependent on the amount of Si present in the core and whether an FeS layer is present. Whether Mercury formed with this composition \textit{in-situ} or began with a nearly solar Fe/Mg and became Fe-rich due to a giant, mantle-stripping collision, is unknown \citep[][and references therin]{Ebel19}. Recently, planets such as K2-229b have been described to have a density of 8.9 $\pm$ 2.1 g/cm$^3$, or a CMF of 68\% \citep{santerne18}, comparable to that of Mercury, but with significant uncertainties that make it also consistent with a nominally ``Earth-like'' structure. It should be noted Mercury's bulk Fe/Mg lies completely outside the the observed range of FGK stars (Figures \ref{fig:Tern} and \ref{fig:FeMg_hist}), suggesting that \textit{if} stellar composition is indeed indicative of rocky planet composition and the Hypatia catalog is a rough approximation of the potential range of stellar Fe/Mg, exo-Mercuries must be either rare and most likely to be found about Fe-rich and $\alpha$-element poor stars. Otherwise, these observed planetary composition must reflect late-stage mantle spallation or another as of yet understood process that occurred during the protoplanetary disk phase. So while large Fe/Mg fractions are possible (we need only look to Mercury), we adopt what we consider to be a conservative range of Fe/Mg for our models, while exploring a relaxation of this constraint in Section \ref{sec:maxP}.

\subsection{Planetary Structure Calculations}
ExoPlex calculates the radius of a planet of specified composition by iteratively solving five coupled differential equations:  the mass within a sphere, 
\begin{equation}
\frac{dm(r)}{dr} = 4\pi r^{2} \rho(r)
\label{eq:MR}
\end{equation}
the equation of hydrostatic equilibrium,
\begin{equation}
\frac{dP(r)}{dr} = -g(r)\rho(r),
\label{eq:he}
\end{equation}
the adiabatic temperature profile,
\begin{equation}
\label{eq:temp}
\frac{dT(r)}{dr} = \frac{\alpha(P, T) g(r)}{C_{P} (P, T)},
\end{equation}
Gauss's law of gravity in one dimension,
\begin{equation}
\frac{1}{r^2}\left(r^{2}\frac{dg(r)}{dr}\right) = 4\pi G \rho(r),
\label{eq:gau}
\end{equation}
and the thermally-dependent equation of state for the constituent minerals,
\begin{equation}
\rho(r) = f(P(r),T(r))
\label{eq:eos}
\end{equation}
where $r$ is the radius, $m(r)$ is the mass , $\rho(r)$ is the density, $P(r)$ is the pressure, $g(r)$ is the acceleration due to gravity, $G$ is the gravitational constant, $T(r)$ is the temperature within a shell of radius $r+dr$, and $\alpha(P, T)$ and $C_{P}(P,T)$ are the thermal expansivity and coefficient of specific heat at constant pressure of the constituent minerals at a given pressure and temperature, respectively. We adopt surface boundary conditions of $P(R) = 1$ bar where $R$ is the final radius of the planet, $g(0)$ = 0 and $m(0)$ = 0. Lastly we adopt $T(R)=T_{Pot}$, where $T_{Pot}$ is the potential temperature: the temperature of the mantle if it were adiabatically decompressed. In reality, a colder, conductive layer is likely present at the surface of planets, transitioning to an adiabat below a surface boundary layer. Effects of temperature on the calculated radius are minor \citep{Dorn15,Unte16}, and thus we first run calculations assuming a single mantle potential temperature,  $T(M)$ = 1600 K, relaxing this constraint in section \ref{sec:PTLim}. 

ExoPlex calculates the stable mantle mineral assemblage for two pressure and temperature grids using PerPlex \citep{Conn09}: a fine mesh grid with 80 temperature and 80 pressure steps encompassing 1 bar/1400 K to 140 GPa/3500 K and a coarse mesh above 100 GPa with 40 temperature and 40 pressure steps from 125 GPa/2200 K to 2.8 TPa/7000 K. This two grid approach captures the behavior of low-pressure phase transitions while reducing phase equilibria calculation time for the relatively simple lower mantle. We adopt the thermodynamic database of \citet{Stix11} for mantle phase equlibria calculations. For the silicate mantle,  $\rho$(P,T), $\alpha$(P,T) and $C_{P}(P,T)$ are calculated as described below and then linearly interpolated from within these P-T-composition grids when solving equations \ref{eq:MR}-\ref{eq:eos}. 

\begin{figure}[t!]
    \centering
    \includegraphics[width=\linewidth]{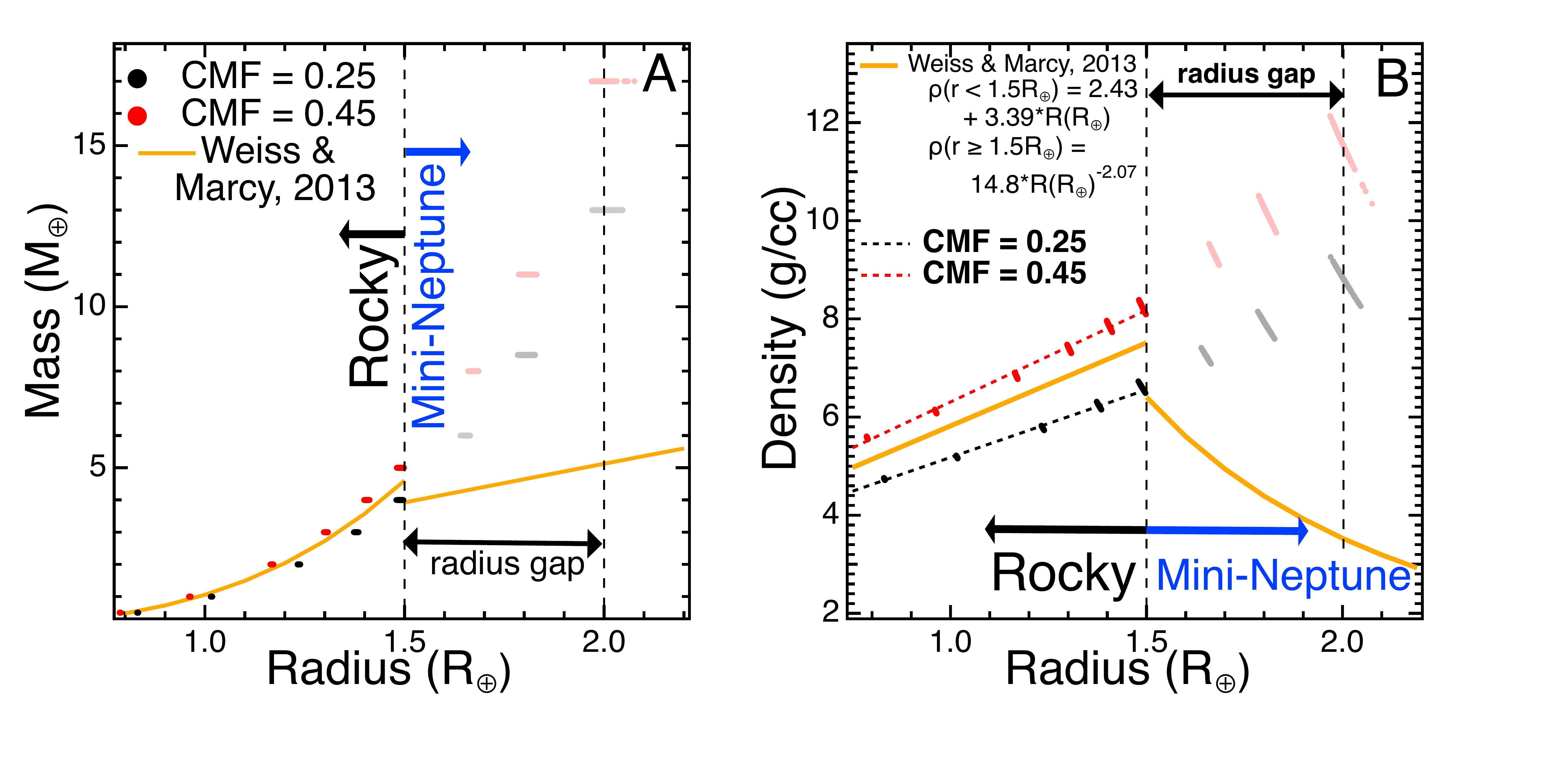} 
    \caption{\textbf{A: } Rocky planet mass as a function of radius for two end-member core mass fraction planets: 0.25 (black; Fe/Mg=0.6) and 0.45 (red; Fe/Mg=1.5). Opaque and transparent points represent 1000 and 200 iterations, respectively, of variation of the mean EoS values about their respective uncertainties. \textbf{B:}  Density (Mass/Volume) as a function of planetary radius for the same mass-fractions. Best fits for radii less than 1.5 R$_{\oplus}$ are shown for CMF = 0.25 (black, $\rho \rm{[g/cc]}=2.42+2.76*R(\rm{R_\oplus}$) and 0.45 (red, $\rho \rm{[g/cc]}=2.57+3.74*R(\rm{R_\oplus}$) are shown as dashed lines. For comparison, we include the density trend derived from binned average of 65 observed exoplanets \citep[yellow, ][]{Weis14}. }

    \label{fig:MR}
\end{figure}

\subsection{Equations of State}

The density of the planetary material in equations \ref{eq:MR}, \ref{eq:he}, \ref{eq:gau} and \ref{eq:eos} is a material property and a function of pressure and temperature. Measurement or calculation of the density at fixed pressures and temperatures are modeled as an equation of state (EoS) in which thermodynamic parameters, including the bulk modulus, $K$, thermal expansion, $\alpha$, and specific heat, $C_P$, can be determined. These thermodynamic parameters are therefore model dependent.  Uncertainties in the fit parameters are therefore a function of the reliability of the pressure standard, the data quality, the to degree to which the form of the EoS effectively models the $P(\rho, T)$ relationship, and the compression range of the data.  

We adopt the recently published isentropic $P-V$ data to 1.4 TPa for pure iron along an isentrope reaching 3000-4000 K at 1.4 TPa \citep{Smith18}. We refit the data to a Vinet EoS along this isentrope incorporating uncertainties in the pressure and density at each measurement ($V_0$=6.662$\pm$0.018 cm$^3$/mol, $K_0$= 175.8$\pm$3.8 GPa, $K'$= 5.64$\pm$0.04). In our fit of the data, we permit uncertainty in the zero-pressure volume of $\epsilon$-Fe along this adiabat, which has the effect of increasing the uncertainty in $K_0$ by a factor of 6, and increasing the uncertainty in $K'$ by a factor of 4, for a conservative estimate of uncertainty within the pressure range of the experiment.

The EoS of $\epsilon$-iron and silicate minerals require an extrapolation of the inferred density beyond the pressure range of the experiments from which the model was derived. Therefore, we model the consequences of the extrapolation through a Monte Carlo approach in which each elastic ($K_{0}$, $K'$, $V_{0}$) and thermal ($\theta_{0}$, $\gamma_0$, $q_{0}$ and $\eta_0$) parameter are randomly determined for $\epsilon$-Fe and each \textit{individual} mantle mineral present in our $P-T$-composition grids. We randomly vary these parameters 1000 times and assume a Gaussian distribution of width equal to the uncertainty in each EoS parameter. Those planets with radii $>$ 1.5 $R_\oplus$ were only run 200 times. Uncertainties in mantle mineral parameters were taken directly from \citet{Stix11}. 
For each run, $\rho(P,T)$, $\alpha(P,T)$ and $C_{P}(P,T)$ are each recalculated for the equilibrium composition reported at every $P$,$T$ point within the PerPlex-derived upper- and lower-mantle grids using the BurnMan software package \citep{Cott14}. We follow the same methodology for randomly determining the EoS parameters of $\epsilon$-Fe in each run adopting the average values and uncertainties for the isothermal EoS derived above and using BurnMan to determine $\rho(P, 300 K)$ within the core. We assume the reported uncertainties in each equation of state parameter to be independent of one another.

This Monte-Carlo approach provides an estimate of the uncertainty in exoplanetary $M$ and $R$ arising from the inherent uncertainties in and extrapolation of the EoS for both mantle and core materials. This approach is valid assuming no change in the nature of compression of a given mineral and no unknown phase transformation. No previous mass-radius forward \citep{Seag07,Unte16,Zeng13} or inverse \citep{Dorn15} models have  adopted any extrapolation of uncertainty in the underlying EoS parameters, instead only adopting the average derived values.

\section{Results}

\subsection{Planet Mass, Radius and Density}
\label{sec:MRD}
We find that the radius of a rocky planet increases with increasing mass and is a strong function of CMF. We find that increasing CMF from 0.25 to 0.45 increases the mass of a 1.5 $R_{\oplus}$ planet by 25\%, from 4 to 5 $M_{\oplus}$ (Figure \ref{fig:MR}, A). Our calculated masses above 1.5 $R_{\oplus}$ are greater than the trends determined by \citet{Weis14} due to the lack of the inclusion of an atmospheric layer in our models. These points above 1.5 $R_{\oplus}$ (Figure \ref{fig:MR}A) then represent compositions indicative of super-Earths, supporting the statistical work of \citet{Weis14}, who predict planets above 1.5 $R_{\oplus}$ to be more likely to be lower mass mini-Neptunes (Figure \ref{fig:MR}A, yellow curve). 

\begin{figure}[t!]
    \centering
    \includegraphics[width=\linewidth]{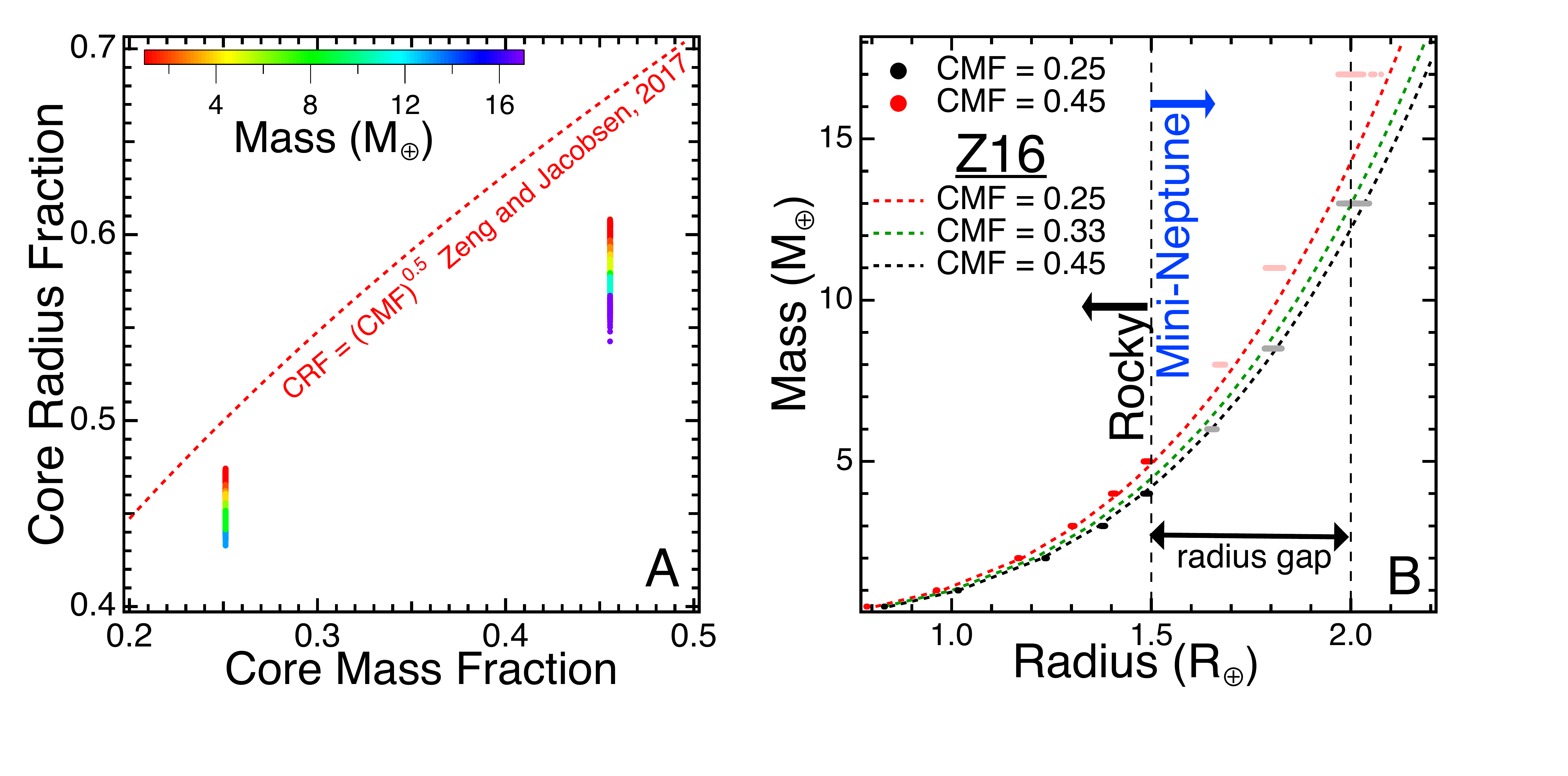}
    \caption{\textbf{A: }Core radius fraction as a function of core mass fraction (0.25 and 0.45). Colors represent the assumed mass of the planet in each run. The empirical scaling relation of CRF $\approx$ CMF$^{0.5}$ proposed in \citet{Zeng17} is shown as a red dashed line. \textbf{B: } Comparison of our calculated mass and radius values (symbols as in Figure \ref{fig:MR}) to the \citet{Zeng16} (Z16) model for CMF = 0.25 (red dashed line), 0.33 (dashed green line) and 0.45 (dashed black line)}
    \label{fig:CMFCRF}
\end{figure}

We calculate the greatest density for likely rocky planets ($R\le$ 1.5 $R_{\oplus}$) to be between $\sim$6.6 g/cm$^3$ and $\sim$8.2 g/cm$^3$ for CMF = 0.25 and CMF = 0.45, respectively (Figure \ref{fig:MR}, B), bracketing the \citep{Weis14} densities for small planets. The CMF range of 0.25 to 0.45 predicts a 25\% difference in total mass at a given radius ($R$ $\le$ 1.5 $R_\oplus$), which is observationally resolvable. Therefore  inferring an exoplanet's CMF is an attainable model result within the uncertainties of the observations for planets with radii less than 1.5 $R_{\oplus}$, offering a key test to the degree to which rocky exoplanets reflect the abundance ratios of refractory elements in the host star. Above 1.5 $R_\oplus$, our modeled density  increases roughly linearly with radius for a volatile-poor super-Earth composition (Figure \ref{fig:MR}B). As with the mass-radius relationships, such $\rho$-$R$ relationships are again in contrast to the exponentially-decreasing observed density trend of \citet{Weis14} (Figure \ref{fig:MR}B, yellow curve) indicative of mini-Neptunes being the likely state of a planet above the transition radius of 1.5 $R_\oplus$.

Despite a planet's mean density and mass sensitivity to CMF, the radius of rocky exoplanets is relatively insensitive to bulk refractory element composition, including Fe/Mg. For example, our models for 4 $M_{\oplus}$ planets vary by just 0.1 $R_{\oplus}$ over the range of CMF modeled. This confirms models of \citet{Unte16} and \citet{Dorn15} that the iron-poor (CMF=0.25) and iron-rich (CMF=0.45) compositions result in just a 6\% difference in radius for rocky planets 4 times the mass of the Earth. We further find the resulting core radius fractions (CRF) varies between $\sim$0.43 - 0.47 for CMF = 0.25 and $\sim$0.55 - 0.58 for CMF = 0.45 (Figure \ref{fig:CMFCRF}). This corresponds to mantle thicknesses of between roughly 4300 and 5500 km, respectively, for a 1.5 $R_\oplus$ planet. 

We also find that an empirical scaling of CRF $\approx$ CMF$^{0.5}$ as proposed by \citet{Zeng17} greatly over predicts the core radius fraction at all masses (Figure \ref{fig:CMFCRF}A). For a 1.5 $R_{\oplus}$ planet, we calculate an average mass 2\% greater and 2\% lower than the mass-radius scaling relationships of \citet{Zeng16} (Z16), for planets with  CMF = 0.25 to 0.45, respectively. Conversely, a 1.5 $R{_\oplus}$ planet of 4 and 5 $M_{\oplus}$, respectively, Z16 predicts a CMF 1\% less than our more exact models. These differences are not resolvable observationally. However, the differences between models increases in the range 1.5-2.0 $R_{\oplus}$, within the radius gap (Figure \ref{fig:CMFCRF} B). In this region of observational space, the Z16 models underpredict mass at a given radius by as much as 16\%. Therefore, use of this simplified model in this radius range will risk misidentifying gaseous planets as rocky planets.

\subsection{Pressure and Temperature Limits}
\label{sec:PTLim}
\begin{figure}[t]
    \centering
    \includegraphics[width=0.9\linewidth]{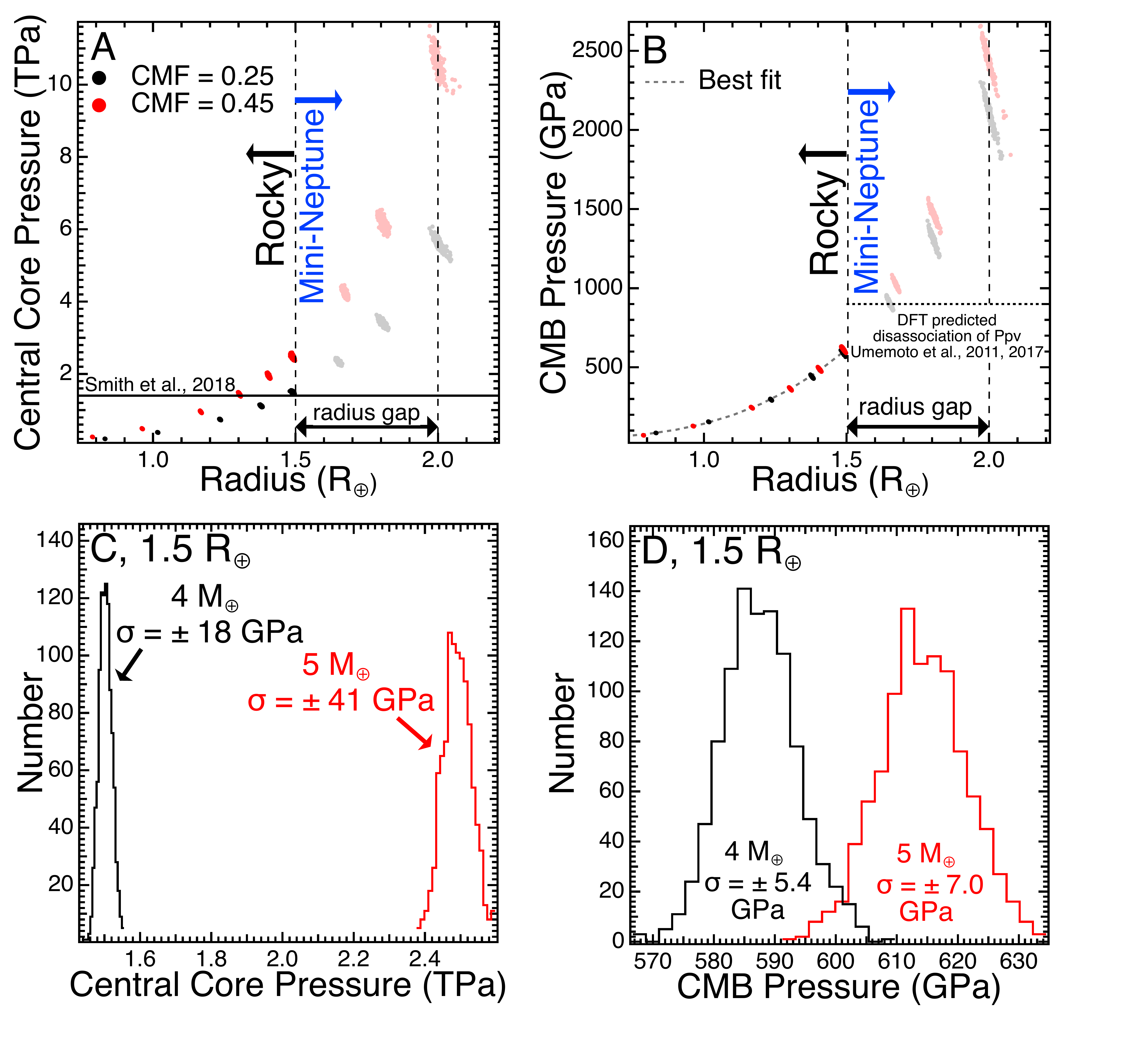}
    \caption{Central core (\textbf{A}) and CMB (\textbf{B}) pressure as a function of radius and for our modeled exoplanets. Symbols as in Figure \ref{fig:MR}. The ``radius-gap,'' where planets less than 1.5 $R_{\oplus}$  are more likely to be rocky planets is shown as a dashed line. We include the highest pressure attained from recent measurement of $\epsilon$-Fe from \citet{Smith18} as a black line in \textbf{A}. \textbf{C \& D} Resulting CMB pressure histograms of 1000 samples of thermoelastic parameters for planets with $\sim$1.5 $R_{\oplus}$ for same CMF as \textbf{A} and \textbf{B}.}
    \label{fig:Pcen}
\end{figure}

The central pressure is similarly sensitive to CMF. This is as a result of CMF being the dominant control of planetary radius at a given mass (Figure \ref{fig:Pcen}, A). For 1.5 $R_{\oplus}$ planets, the central pressure increases from 1.5 TPa when CMF = 0.25 to 2.5 TPa when CMF = 0.45 (Figure \ref{fig:Pcen}: C). The central pressure for a 1.5 $R_{\oplus}$ with CMF = 0.25 is $\sim$100 GPa greater than the currently greatest pressures reached in the shock experiments of \citet{Smith18}. The central pressures of the more iron-rich planets, however, require a nearly 100\% extrapolation in pressure. Despite these high pressures, the range of calculated central core pressures due to uncertainties in the EoS varies by 18 and 41 GPa for CMF = 0.25 and 0.45, respectively, or less than $\sim$2\% (Figure \ref{fig:Pcen}C). 

The pressure at a rocky planet's core mantle boundary (CMB) also increases with increasing planet radius reaching a maximum of $\sim$630 GPa at 1.5 $R_{\oplus}$ (Figure \ref{fig:Pcen}: B) with the largest uncertainties in this value being due to uncertainty on the order of 1\% in the EoS (Figure \ref{fig:Pcen}: D). This maximum CMB pressure for likely rocky super-Earth exoplanets is more than twice the greatest room temperature experimental measurement of Ppv, but below the predicted pressure of the breakdown of Ppv to constituent oxides \citep{Umem11}. CMB pressure does increase above 1.5 $R_{\oplus}$ for a rocky, super-Earth composition; however, these planets are unlikely to be super-Earths, but mini-Neptunes with a significant volatile surface layer \citep{Weis14}. 

In contrast to planetary radius and central pressure, we find a planet's CMB pressure is not a strong function of CMF. While a 1.5 $R_{\oplus}$, CMF = 0.45 planet has a mantle 87\% the thickness of the mantle of 1.5 $R_{\oplus}$, CMF = 0.25 planet, the greater CMF leads to an increased $g(r)$ due to larger core (equation \ref{eq:he}). This effect increases the hydrostatic pressure gradient, which nearly balances the effect of the high CMF planet having a shallower overlying mantle compared to the low CMF planet. For instance, a 1.5 $R_{\oplus}$, 4.0 $M_{\oplus}$ planet (CMF = 0.25), has a CMB pressure that is just 30 GPa less than a 5.0 $M_{\oplus}$ planet of the same radius, but with larger core mass fraction (CMF = 0.45), a difference less than the typical accuracy of experimental pressure measurements at that pressure (Figure \ref{fig:Pcen}D). Therefore, the pressure at a planet's CMB is primarily a function of total planet radius, in which the pressure \textit{gradient} in the mantle increases with increasing core fraction. Fitting our calculations for planets with $0.75 R_{\oplus} \leq R \leq 1.5 R_{\oplus}$ a best fit to our modeled CMB pressures is: 
\begin{equation}
P_{\rm CMB} \rm{(GPa)} = 262\textit{R} - 550\textit{R}^2 + 432\textit{R}^3 
\label{eq:CMBP}
\end{equation}
where $R$ is expressed in Earth radii. For a 1 Earth radius planet, equation \ref{eq:CMBP} yields a CMB pressure of 144 GPa, overestimating the actual pressure of 136 GPa \citep{PREM} by 6\%, a consequence of modeling the core as pure, solid iron (see \ref{sec:maxP}).

\begin{figure}[t!]
    \centering
    \includegraphics[width=\linewidth]{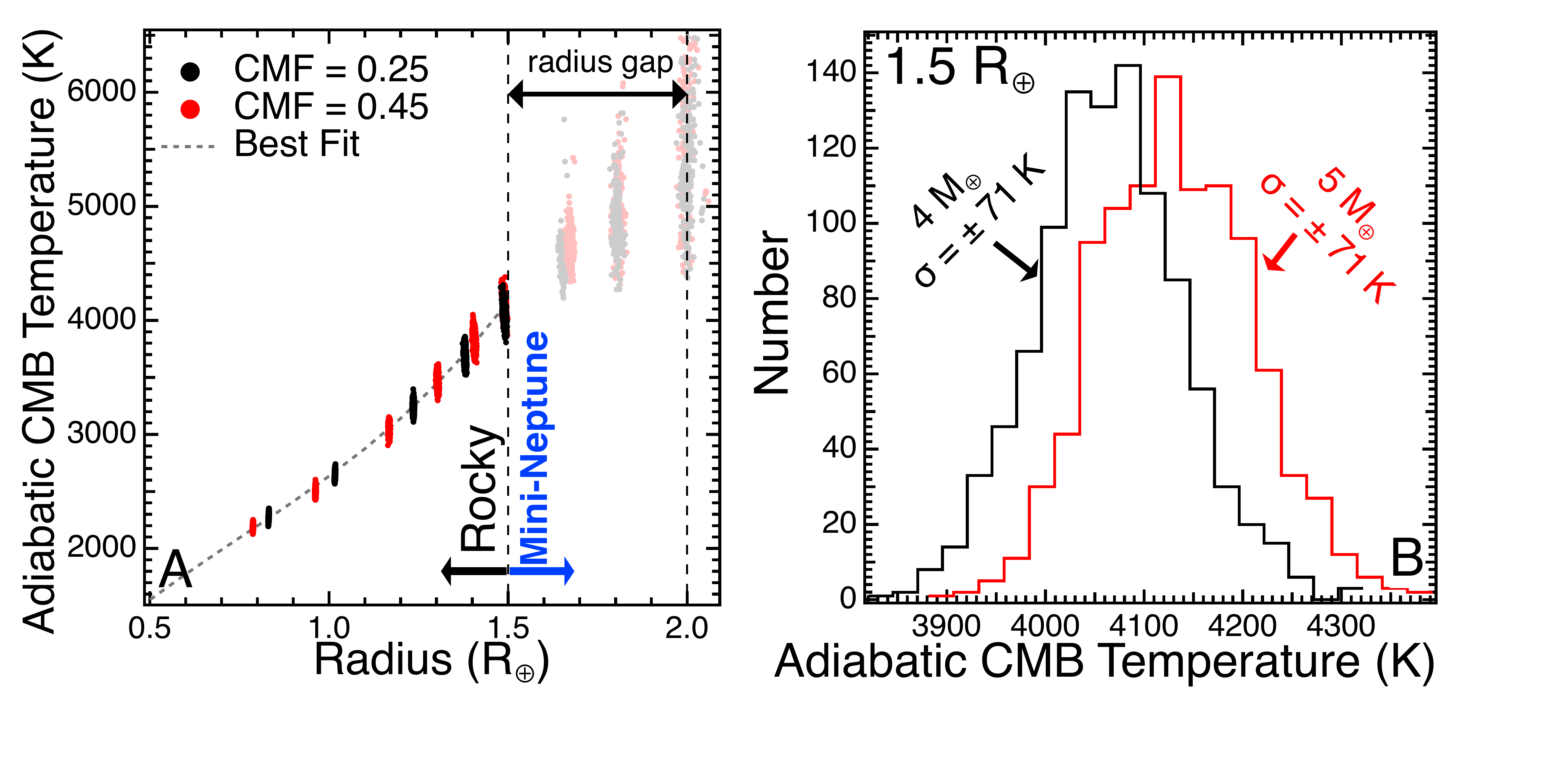}
    \caption{\textbf{A}: Core-mantle boundary temperature calculated along adiabats for two end-member core mass fraction planets assuming a 1600 K mantle potential temperature. Symbols as in Figure \ref{fig:MR}. \textbf{B:} Resulting CMB temperature histograms of for planets $\sim$1.5 $R_{\oplus}$.}
    \label{fig:CMBT}
\end{figure}

Finally, we constrain the adiabatic temperature gradient through the silicate mantle. Assuming an Earth-like mantle potential temperature of 1600 K (Figure \ref{fig:CMBT}), we find CMB temperature increases with planetary radius reaching a maximum of $T\sim4100$ K at 1.5 $R_{\oplus}$ planets. As with the CMB pressure, the adiabatic temperature gradient is primarily a function of the hydrostatic pressure gradient (equation \ref{eq:he}), and therefore insensitive to the bulk planetary structure as reflected in the core mass fraction. Fitting our calculations for planets with $0.75 \leq R \leq 1.5 R_{\oplus}$, we find a best fit for our modeled CMB temperature when mantle potential temperature is 1600 K to be:

\begin{equation}
T_{\rm CMB} \rm{(R)} = 4180\textit{R} - 2764\textit{R}^2 + 1219\textit{R}^3 
\label{eq:TCMB}
\end{equation}

\noindent where $R$ is expressed in Earth radii. Equation \ref{eq:TCMB} yields a CMB temperature of 2635 K for a 1 Earth radius planet, in good agreement with the 2500-2800 K Earth value of \citet{Lay08} as determined using a similar method to ours. Above 1.7 $R_{\oplus}$, the PerPlex-derived phase diagrams predict the disassociation of Ppv into component oxides (MgO and SiO$_2$). The large uncertainty of the SiO$_2$ value for volume exponenet of the grŸneisen parameter, $q_{0}$, reported in \citet{Stix11}  (stishovite: $q_{0} = 2.8 \pm 2.2$) leads to a wide range of calculated $\alpha (P,T)$ and $C_{P} (P, T)$ in equation \ref{eq:temp}. As such, the range of predicted adiabatic CMB temperatures increases above 1.7 $R_{\oplus}$, while conclusions based primarily on compression (planetary radius, CMB pressure, central core pressure) vary only moderately (Figures \ref{fig:MR} and \ref{fig:Pcen}).

\begin{figure}[t!]
    \centering
    \includegraphics[width=0.75\linewidth]{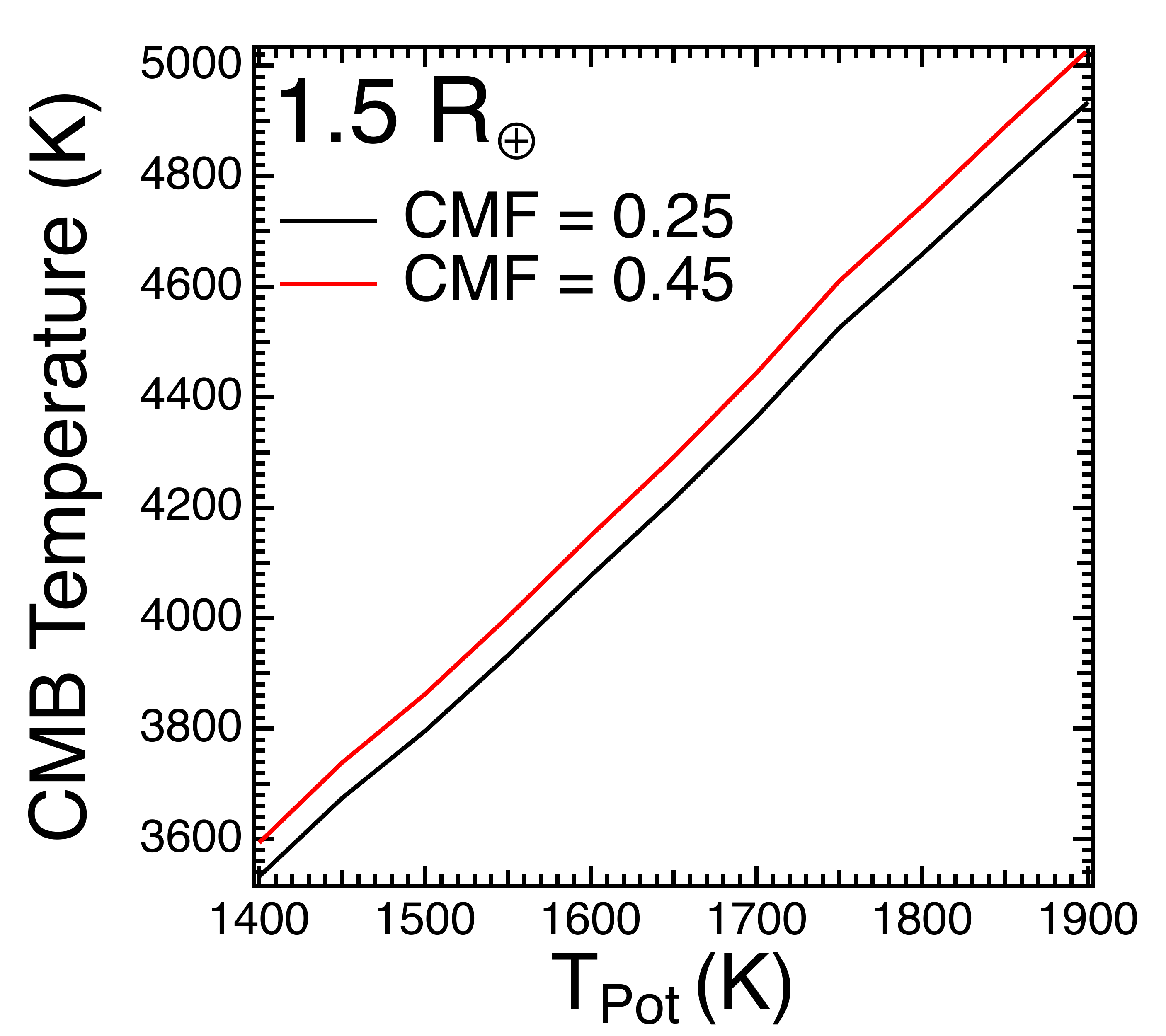}
    \caption{CMB temperature for a 1.5 $R_{\oplus}$ planet (CMF=0.25 black, CMF=0.45 red) as a function of mantle potential temperature ($T_{Pot}$) using mean EoS parameters. }

    \label{fig:Tpot}
\end{figure}

Relaxing the constraint of an Earth-like mantle potential temperature, we further consider planets with potential temperatures of 1400 K and 1900 K. As compositions of $0.5 \le \rm{Mg/Si} \le 1.2$ in the Mg$_2$SiO$_4$-SiO$_2$ binary will be liquid at the surface at 1900 K potential temperature. Thus, considering hotter adiabatic temperature profiles would not be representative of a solid planet, but one marked by surface magma oceans. These models adopting this range of $T_{Pot}$ broaden the temperature range of CMB temperatures in 1.5 $R_{\oplus}$ planets from $\sim$3500 K to 5000 K (Figure \ref{fig:Tpot}). On the other hand, the difference in CMB temperature of a 1.5 $R_{\oplus}$ planet between the end-member CMF  models differ from each other by $\sim$50 K for a given mantle potential temperature (Figure \ref{fig:Tpot}). The change in adiabatic temperature model that is a function of both planetary radius ($R$) and potential temperature ($T_{Pot}$) is:
\begin{equation}
    \Delta T_{CMB}(R, T_{Pot}) \approx (T_{Pot} - 1600\ \rm{K})*(0.82 + R^{1.81})
\label{eq:delta}
\end{equation}
where $R$ is expressed in Earth radii and is again valid between 0.75 and 1.5 R$_\oplus$. Combining equations \ref{eq:TCMB} and \ref{eq:delta} yields the total change in adiabatic CMB temperature as a function of planet radius and mantle potential temperature. As in the 1600 K case (Equation \ref{eq:TCMB}), this temperature is roughly independent of CMF and therefore this model predicts that the change in adiabatic temperature \textit{gradient} is primarily a function of CMF alone, regardless of planet radius. Tidal heating and thermal boundary layers within the mantle can all lead to significantly super-adiabatic temperature profiles. 

\subsection{Effect of Uncertainties in Equations of State}
The impact of EoS uncertainties is minor due to the extrapolation of the EoS for $\epsilon$-Fe and Ppv, the two dominant interior phases that require significant extrapolation in super-Earth's. To the extent that neither material undergoes a transition such that the character of compression changes, uncertainties in the EoS of each account for $\sim$0.25\% uncertainty in radius for a 4 $M_{\oplus}$ planet.  These uncertainties are less than the effect of variable CMF on radius for a given mass, as well as below the observational resolution for all but well-resolved timed transit variation (TTV) studies \citep[e.g.][]{Grimm18}. Similarly, the uncertainties in the EoS lead to just $<$2\% uncertainty for central and CMB pressures (Figure \ref{fig:Pcen}: B, D), and CMB temperature (Figure \ref{fig:CMBT}: B). These uncertainties, reflecting just the extrapolation of the model fit to the compression data, likely overstate the exactness of the interior models, but illustrate that increasing the pressure range of compression measurements will not affect the inferred CMF, or presence, or lack of, a significant gaseous layer. 

\section{Discussion}
Mass and radius currently provide our only direct observables for inferring the interior structure and composition of rocky exoplanets. In addition to interior mineralogy and relative core mass, mass-radius-composition models also yield the temperature and pressure range of each layer, important for modeling interior dynamics. Adiabatic gradients are important in determining the depth of surface melting and convective vigor \citep[e.g.][]{Kite09,Noack17,Dorn18}, while pressure gradients affect the depth to solid-state phase transitions. Absolute pressure and temperature, in turn, outline the parameter space needed to be explored by future experiments and ab-initio models for determining the physical properties of super-Earth mantles and cores.  As a result, significant effort has been given to constraining both metallic Fe and silicate mineral phase transitions in the TPa regime. Here we empirically show that the primary variable in mean density of a rocky exoplanet, the core mass fraction, has little effect on the pressure and temperature limits of a silicate mantle, reducing the problem to a function of the planet's radius.

\begin{figure}[t!]
    \centering
    \includegraphics[width=0.8\linewidth]{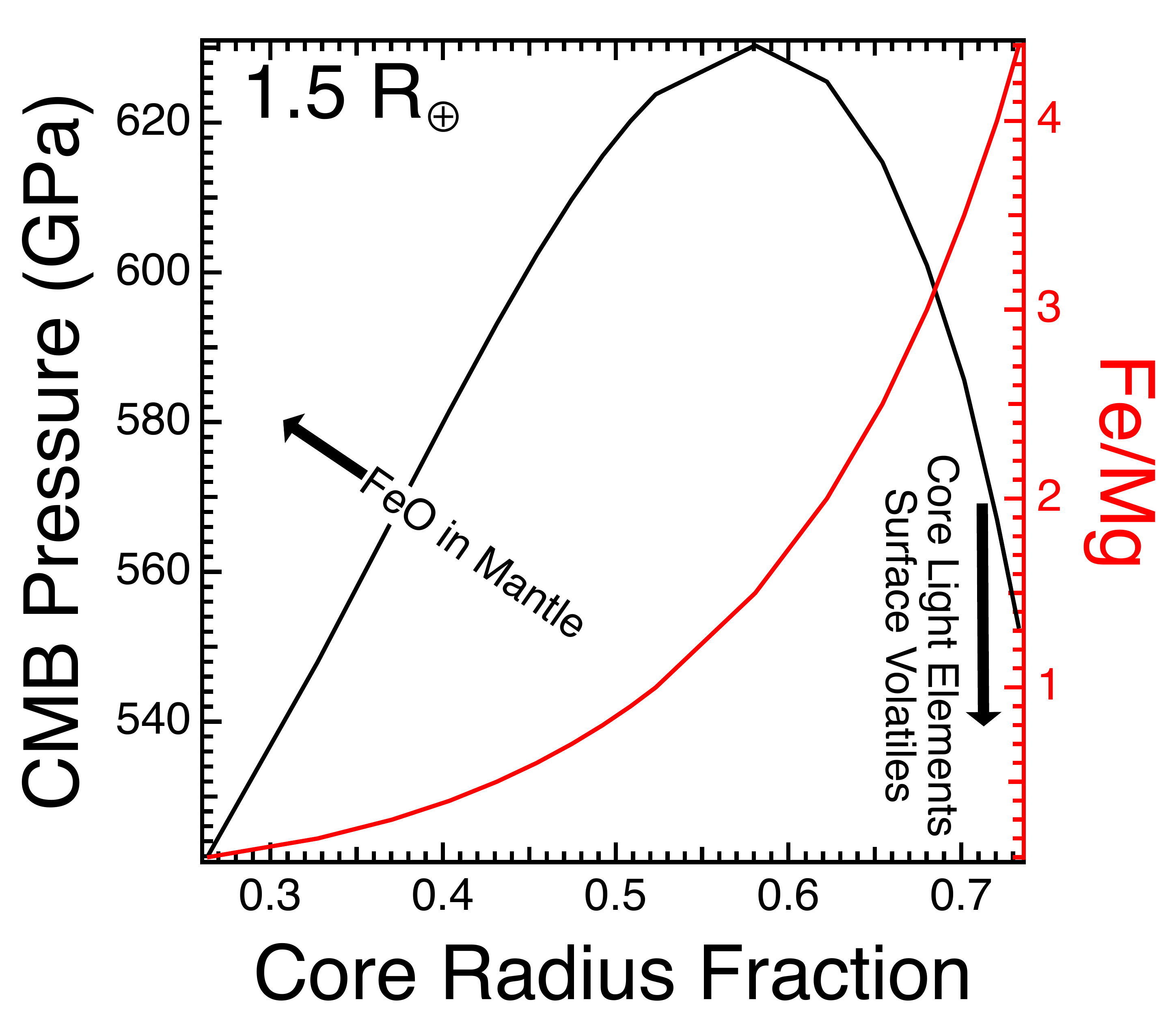}
    \caption{Model results of core-mantle boundary pressure (black) and input bulk planet Fe/Mg (red) as a function of core radius fraction for a 1.5 $R_{\oplus}$ planet with mantle Si/Mg = 1. A core radius fraction of 0.45 and 0.58 are derived for planets with Fe/Mg = 0.6 and 1.5, respectively.}
    \label{fig:FeMg}
\end{figure}

\subsection{Upper CMB Limits for Rocky super-Earths Exoplanets below 1.5 $R_{\oplus}$}
\label{sec:maxP}
Rocky planets that reflect their refractory stellar abundances and lack a significant volatile-rich envelope with $R \leq 1.5$ $R_\oplus$ are limited in mass to 5 Earth masses for Fe/Mg $<$ 1.5 and Si/Mg $\sim$ 1. Changing Si/Mg relative to Fe/Mg and variable mantle potential temperature will both have minimal impact on the mantle density within observational uncertainties, typically 20\% in mass and 4\% in radius. \citep{Dorn15, Unte16}. We calculate that for planets smaller than 1.5 Earth radii, the core-mantle boundary pressure is, to first order, a function planetary radius (equation \ref{eq:CMBP}), reaching a maximum pressure of 630 GPa at 1.5 $R_{\oplus}$. This radius represents the transition at which point planets are more likely to be rock-iron dominated super-Earth's to volatile-rich mini-Neptunes. This CMB pressure is then a likely maximum for those planets most likely to be rocky super-Earths (R $<$ 1.5 R$_\oplus$) as opposed to mini-Neptunes (R $>$ 1.5 R$_\oplus$) with respect to core mass fraction, core composition, and presence of a volatile-rich surface layer. 

For $R \leq 1.5$ $R_\oplus$ planets, a wider range of Fe/Mg from Fe-Free (Fe/Mg = 0) to Mercury-like (minimum Fe/Mg = 3.9) results in CRF ranging from 0 to 0.64 (Figure \ref{fig:FeMg}). The resulting masses and central pressures as a consequences of this broader range of Fe/Mg then reflect a wider range than considered for our modeling above. Increasing Fe/Mg increases the mass of the planet, central pressure, and mantle pressure gradient. However, increased CRF also leads to a decrease in the thickness and mass of the overlying mantle (Figure \ref{fig:CMFCRF}), causing CMB pressure to decrease at high CRF. The trade off is such that the maximum CMB pressure occurs at a CRF = 0.55 (CMF = 0.45, Fe/Mg = 1.5; Figure \ref{fig:FeMg}).  Exo-Mercuries with Fe/Mg $>$ 1.5 are possible and the mass of such planets will increase with Fe/Mg (and thus CRF) at constant radius. Remarkably, however, our central conclusions of maximum CMB pressure and geothermal temperature ranges are largely unaffected, with iron-rich exo-Mercuries having lower values for each due to this trade off between mass of overlying mantle and pressure gradient at the CMB (Figure \ref{fig:FeMg}). 

Planets with stellar refractory element abundances (Fe, Mg, Si) when oxidized such that significant iron is retained in the mantle \citep[e.g.][]{ET08,Schaefer17} will have smaller cores than assumed here. Given constant bulk Fe/Mg, as mantle FeO content increases, a planet's CMF decreases. As a consequence, the depth of the core mantle boundary increases due to a smaller core radius. As a result, a Fe/Mg = 1.5 planet in which 20\% of the Fe is stored in the mantle, the CMF reduces from 45\% to 35.6\%. This effect leads to decreased planetary mass and surface gravity (equation \ref{eq:gau}) at a constant planet radius. Pressures do increase moderately due to increased mantle density, mitigating the trade off between a deeper CMB and decreased planet mass (Equation \ref{eq:he}). In the case of 20\% Fe stored in the mantle as FeO \citep[as in Mars, ][]{wanke94}, a 1.5 $R_\oplus$ planet with bulk Fe/Mg = 1.5 has a CMB pressure of 740 GPa. Mantle oxidation to Fe$_2$O$_3$ or incorporation of oxygen into the core will mitigate this effect. Extreme oxidation of the planet is the only factor we identify as increasing the CMB pressure beyond our predicted maximum CMB pressure of 630 GPa for a 1.5 $R_\oplus$ planet. These higher pressures, then, are only valid for those planets with silicate compositions that contain FeO.

Planetary cores are not pure, solid iron, but likely mostly molten with a fraction of light elements. Our rocky exoplanetary modeling approach assuming pure solid iron cores leads to an upper bound in CMB pressure at a given Fe/Mg and Si/Mg. Those cores composed of liquid iron and/or containing light elements will have a lower density relative to $\epsilon$-Fe used in our models  \citep{Unte16,Schaefer17}. In the case of a liquid iron core without light elements, the CRF of a planet is negligibly larger compared to our models adopting a solid $\epsilon$-Fe core \citep{Unte16}. However, in a planet with a core containing light elements, but with constant core mass fraction, there will be an overall reduction in planet mass. The reduced core density thereby decreases the pressure at the CMB due to the reduced mass of the planet. Similarly, the introduction of light elements to the core, at constant Fe/Mg and Si/Mg, reduces the mean molecular weight of the core and the planet, again reducing the mass of the planet, with associated reduced pressure gradients within the mantle.

Finally, while not meeting our initial definition of a two-layer rocky planet, for a given planet radius, the inclusion of a volatile layer on the surface of a planet (e.g. H$_2$/He atmosphere or water layer) will also reduce CMB pressure at fixed size by lowering the density of the overlying material at the CMB. 

Therefore, we assert that the pure-Fe core and Fe-free silicate mantle case modeled here represents a useful upper-limit range of pressures likely to occur in the rocky planets below the transitional planet radius of $R \leq 1.5$ $R_\oplus$ where planets are observationally determined to be more likely to be rocky super-Earths rather than volatile-rich mini-Neptunes. While CMB pressures may increase above 630 GPa due to changes in the oxidation state of Fe, these high pressures are only possible in those silicate mantle compositions with large fractions of FeO. 

Our calculated upper limit for CMB pressure for likely rocky planets is about 100 GPa greater the expected MgO+MgSiO$_3$ or SiO$_2$+MgSiO$_3$ recombination reactions \citep{Umem17}, but well below the predicted pressure at which post-perovskite is expected to dissociate into oxide components \citep[Figure \ref{fig:Pcen}B, ][]{Umem11, Umem17}.  While these phases may exist in planetary interiors, they are more likely to occur in the larger mini-Neptunes ($R \geq 1.5 R_{\oplus}$) with extended gas envelopes rather than rocky super-Earths. When focused instead on exoplanetary dynamics, recent work by \citet{Berg18} examined the effects of increasing mass on mantle dynamics, however, these models were performed for purely rocky planets up to 20 $M_{\oplus}$ for a single core-mass fraction based on mass-radius models from \citet{Sot07}. This mass range extends well beyond the mass range predicted from rocky exoplanet occurrence rates ($R$ $<$ 1.5 $R_{\oplus}$). 

\section{Conclusion}
We show that uncertainties in EoS have a negligible effect on extrapolated planet radius and inferred planet properties compared to bulk composition. Thus, while high-pressure experiments to better constrain the EoS of post-perovskite and $\epsilon$-Fe may aid in our understanding of the Earth's core mantle boundary region and core, such measurements are unlikely to do more than fine-tune our density calculations for mass-radius calculations. 

Mass-radius models permit the inference of planetary composition insomuch as we can gauge whether it is mostly rocky or having an extended gas atmosphere. They will not, however, provide many constraints on the interior mineralogy until the precision of mass and radius improve to $\sim$ 1\% \cite{Dorn15,Unte16}. Similarly, both the abundance of light elements in a planet's core and the relative amount of FeO in an exoplanet's mantle lower the planet's density, thus mimicking the signal produced from an extended volatile envelope, limiting our ability to definitively determine the structure of a potentially water-rich exoplanet \citep[][b]{Unte18b}. 

Stellar compositions have been proposed as a proxy for rocky planet composition, with recent work utilizing this to explore the range of potential mineralogies of these planets through forward modeling \citep[e.g. ][]{Hink18}. While planetary mineralogy is an important zeroth-order control of a planet's state and evolution, much is yet to be constrained as to the dynamical and chemical \textit{consequences} of a planet as a result of variable composition. For example, much of our information of melting curves along with composition of coexisting liquids and solids is known for compositions near that of the Earth. Elemental abundances of stars, however, outline a potential compositional parameter space well outside of the Sun and Earth's \citep[Figure \ref{fig:Tern}, ][]{Hink18}.

We show that for planets smaller than 1.5 $R_{\oplus}$ and a likely range of CMF, the likely pressure range over which silicates are present in super-Earths is independent of planetary structure and does not extend beyond $\sim$600 GPa and $\sim$5000 K. We define this as a ``maximum'' for rocky super-Earths as above the transitional radius of 1.5 $R_{\oplus}$, planets are observationally determined to be more likely volatile-rich mini-Neptunes as opposed to rock-dominated super-Earths \citep[Figure \ref{eq:MR}, ][]{Weis14}. We, therefore, propose these values a useful guide and upper-limit for classifying whether experimental and ab-initio results are indicative of super-Earths or mini-Neptunes.  Only for those mantle compositions with significant FeO, do pressures rise above this threshold, and only marginally even for Mars-like FeO contents. Instead, we advocate that to better aid the exoplanet field, experiments and ab-initio calculation should focus on thermal properties, the effects of multi-component systems, melting temperatures and phase relationships, and strength of the materials. These together will provide critical constraints for use in models of the dynamical evolution of such planets. Furthermore, in addition to extending high-pressure solid iron shock-wave work to 2.5 TPa,  experiments and calculations in iron must extend to the physical properties of liquid iron, including compressibility, thermal, and electrical properties at pressures between 0.5-2 TPa.

\acknowledgments
CTU acknowledges the support of Arizona State University through the SESE Exploration fellowship. The results reported herein benefited from collaborations and/ or information exchange within NASA's Nexus for Exoplanet System Science (NExSS) research coordination network sponsored by NASA's Science Mission Directorate. WRP acknowledges the support of NSF EAR 1724693. The unmodified ExoPlex software package is available at https://github.com/CaymanUnterborn/ExoPlex. The Hypatia catalog is available at http://hypatiacatalog.com. \\ 
\noindent The data for all figures is available at: https://tinyurl.com/y7wlvamw. 

\bibliographystyle{agufull08}

\bibliography{agusample}

\end{document}